# Valley Topological Phases in Bilayer Sonic Crystals

Jiuyang Lu,[1] Chunyin Qiu,[2]* Weiyin Deng,[1] Xueqin Huang,[1] Feng Li,[1] Fan Zhang,[3] Shuqi Chen,[4]* and Zhengyou Liu[2,5]*

[1]School of Physics and Optoelectronic Technology, South China University of Technology, Guangzhou, Guangdong 510640, China

[2]Key Laboratory of Artificial Micro- and Nanostructures of Ministry of Education and School of Physics and Technology, Wuhan University, Wuhan 430072, China

[3]Department of Physics, University of Texas at Dallas, Richardson, Texas 75080, USA

[4]Key Laboratory of Weak Light Nonlinear Photonics of Ministry of Education, School of Physics, Nankai University, Tianjin 300071, China

[5]Institute for Advanced Studies, Wuhan University, Wuhan 430072, China

**Abstract:** Recently, the topological physics in artificial crystals for classical waves has become an emerging research area. In this Letter, we propose a unique bilayer design of sonic crystals that are constructed by two layers of coupled hexagonal array of triangular scatterers. Assisted by the additional layer degree of freedom, a rich topological phase diagram is achieved by simply rotating scatterers in both layers. Under a unified theoretical framework, two kinds of valley-projected topological acoustic insulators are distinguished analytically, i.e., the layer-mixed and layer-polarized topological valley Hall phases, respectively. The theory is evidently confirmed by our numerical and experimental observations of the nontrivial edge states that propagate along the interfaces separating different topological phases. Various applications such as sound communications in integrated devices, can be anticipated by the intriguing acoustic edge states enriched by the layer information.

*Corresponding authors.
cyqiu@whu.edu.cn; schen@nankai.edu.cn; zyliu@whu.edu.cn

The discovery of topological insulators, signaled by the presence of symmetry protected edge states, has opened up new avenues for condensed-matter physics [1-4], because of great interest in fundamental physics and prospective applications (e.g., in quantum computing). Recently, intense efforts have been devoted to realizing classical analogues of two-dimensional (2D) topological insulators for photonic, mechanical and sound waves [5-31]. The macroscopic characteristic plus the flexibly tunable crystal symmetry enable these classical systems to be good platforms to investigate the topological properties predicted originally in electronic systems. The topologically protected edge modes could be particularly attractive to overcome some disorder-related restrictions in photonic and acoustic technologies.

Basically, the existing 2D topological insulators for classical waves can be classified into two groups: those mimicking integer quantum Hall insulators with broken time-reversal (TR) symmetry, and those mapping to quantum spin Hall (QSH) insulators with intact TR symmetry. To break the TR symmetry, magneto-optic effects [5-9], gyroscopic metamaterials [10,11], and circulating fluid flows [12-15] have been introduced to the photonic, mechanical and acoustic systems, respectively. Resorting to the paraxial approximation in phase modulated waveguides [16-18], effective gauge flows have also been used to simulate the quantum Hall effect. To design classical QSH insulators, pseudospins must be constructed since the classical waves lack intrinsic half-integer spins. The early attempts have been focused on 2D photonic systems, in which different transverse polarization modes are combined together to mimic the Kramers doublet [19-22]. Similar approaches have been further extended to elastic [23] and mechanical systems [24]. Recently, degenerated Bloch modes induced by high crystalline symmetries have also been proposed to realize pseudospins for polarized light [25,26] and scalar sound [27-29].

In addition to spin, valley degree of freedom has been proved to be another controllable degree of freedom for electrons and recently attracted much interest in 2D layer structures [32-37]. The nontrivial Berry curvature may also contribute topological edge transport without breaking TR symmetry [38-43]. The valley index is easy to migrate to the classical systems by breaking mirror or inversion symmetry [44-47]. This provides another efficient recipe to realize topological edge modes for classical waves [48-54]. Interestingly, Lu *et al.* have observed the valley-projected edge transport of sound in a monolayer sonic crystal (SC) [49], where the topological phase transition is realized by simply rotating the anisotropic scatterers. In this Letter, we propose a new strategy to achieve topological sound transport by designing a bilayer SC (BSC) made of rotated scatterers. Combining the valley and additional layer indices together, the unique bilayer system exhibits a richer topological phase diagram than those explored previously [27-29,49]. Our analytical model demonstrates that the topological phases can be characterized by two quantized topological invariants. The presence of nontrivial acoustic edge modes, either layer-mixed or layer-polarized, has been validated numerically and experimentally. As a manifestation of prospective applications of our bilayer system, an intriguing inter-layer converter has been conceived further for flipping the layer-polarization.

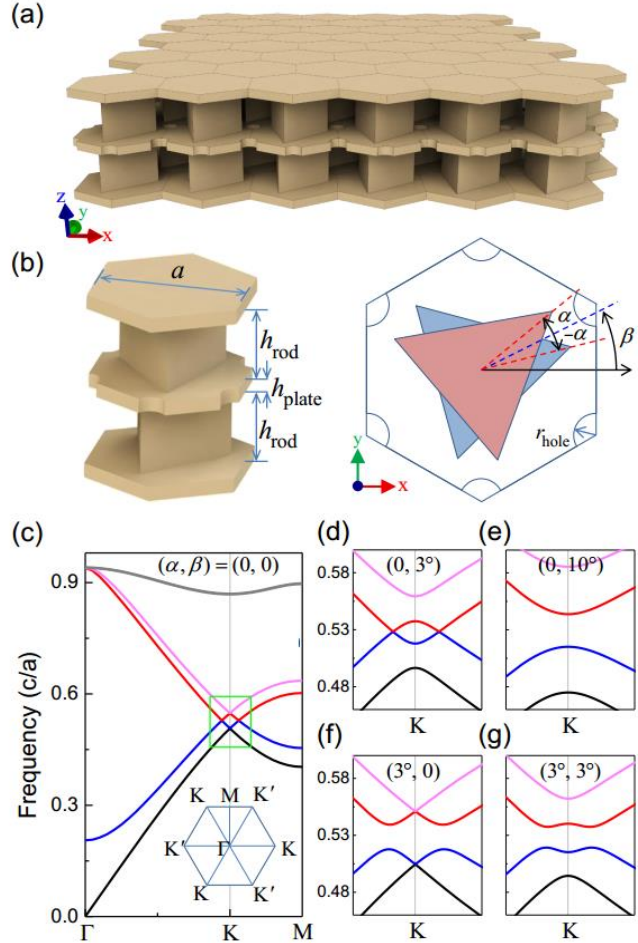

FIG. 1. (a) Schematic of the coupled BSC of hexagonal lattice. Air is filled inside the structured channels with hard boundaries. (b) Side (left) and top (right) views of the unit cell. Specifically, the angles $\alpha$ and $\beta$ together characterize the orientations of the triangular scatterers in both layers. (c) Numerical dispersion for the BSC with zero angles. Inset: the first Brillouin zone. (d)-(g) Local views of the band structures for the four BSCs with specified rod orientations.

As shown in Fig. 1(a), the BSC consists of two layers of SCs sandwiched between a pair of rigid plates and separated by a plate penetrated with a honeycomb array of circular holes. Each layer consists of a hexagonal array of regular triangular rods. Specifically, we consider the following geometrical parameters: the volume filling ratio of the triangular rod $R = 0.24$, the height of the rod $h_{rod} = 0.5a$, the thickness of the middle plate $h_{plate} = 0.1a$, and the radius of the hole $r_{hole} = 0.1a$, with $a$ being the lattice constant. We use the relative angle $\alpha$ and the common angle $\beta$ to characterize the orientations of the triangular rods in both layers [see Fig. 1(b)]. As shown below, the combination of the angles $(\alpha, \beta)$ enables the BSCs with various types of band structures. The simulation is performed by COMSOL Multiphysics based on the finite-element method.

We start from the BSC with $\alpha = \beta = 0$. It has been pointed out that, for a hexagonal monolayer SC with unrotated triangular rods, Dirac degeneracy emerges at the Brillouin zone corners K and K' owing to the protection of $C_{3v}$ symmetry [55]. Here the dispersion of the BSC [Fig. 1(c)] exhibits two conic degeneracies, which are deterministically protected by $D_{3h}$ symmetry that supports two nonequivalent 2D irreducible representations. Interestingly, the 2$^{nd}$ and 3$^{rd}$ bands intersect at an equifrequency ring (i.e., the so-called nodal ring) enclosing the K point. Here band repulsion does not occur since these two bands possess opposite mirror eigenvalues about the middle plane. As the scatterers are rotated, in general, the point and/or ring degeneracy will be broken due to the symmetry reduction [see Figs. 1(d)-1(g)]. To explore acoustic topological insulators, it is crucial to open an omnidirectional gap, which can be realized by breaking the nodal ring degeneracy between the 2$^{nd}$ and 3$^{rd}$ bands. This has already been shown in Fig. 1(e) associated with $\alpha = 0$, where the ring degeneracy is lifted beyond a threshold value of $\beta$, and also shown in Figs. 1(f) and 1(g) associated with $\alpha \neq 0$, which breaks the mirror symmetry with respect to the middle plane.

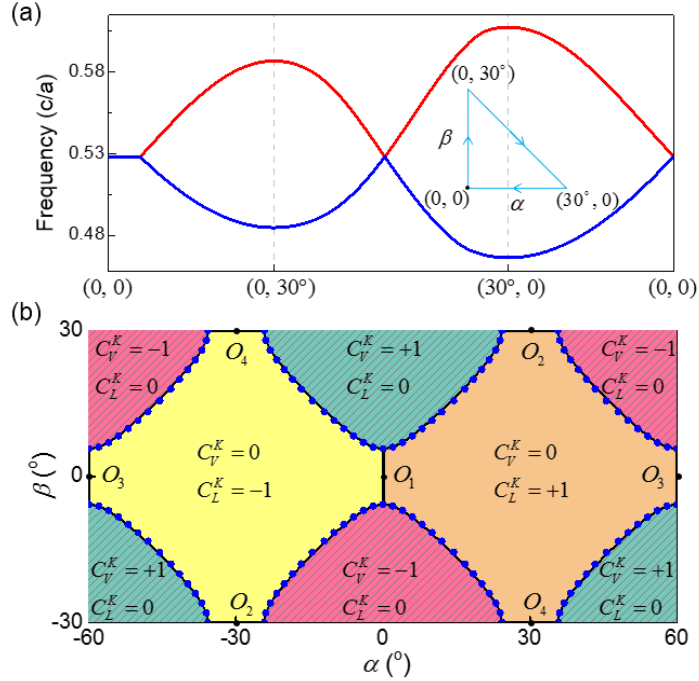

FIG. 2. (a) Band edge frequencies of the 2$^{nd}$ and 3$^{rd}$ bands, varied along the angular path depicted in the inset. (b) Reduced phase diagram parameterized by the angles $(\alpha, \beta)$, distinguished by the quantized topological invariants $C_V^K$ and $C_L^K$. The numerical phase boundaries (solid lines) correspond to the closure of the omnidirectional bandgap, consistent with the model predictions (dots).

To further search the angular boundaries that distinguish different topological phases, in Fig. 2(a) we plot the edge frequencies for the 2$^{nd}$ and 3$^{rd}$ bands along a representative angular path. It is observed that, an omnidirectional bandgap exists in general, except for the line segment from $(0,0)$ to $(0,5.7°)$ and a crossing point at $(6.0°,8.4°)$. The former corresponds to the aforementioned nodal ring degeneracy, whereas the latter stems from an accidental Dirac point degeneracy. The study extended to the whole angular domain gives a reduced phase diagram [Fig. 2(b)], considering the threefold rotation symmetry of the triangular rods. The phase diagram is separated by the straight and curved lines associated with ring and point degeneracies, respectively. Hereafter we focus on the topological properties of the angular regions around the origin point $O_1$, and the phase domains around the special points $O_2$, $O_3$ and $O_4$ can be analyzed similarly.

Starting from the unrotated BSC in the absence of interlayer coupling (Supplemental Materials [56]), an effective Hamiltonian near the K point can be developed to describe the topological phases around the $O_1$ point. (The physics in the K' valley can be derived by considering TR symmetry.) Spanned by the four degenerate states at K point, the perturbation Hamiltonian of the BSC has the form

$$\delta H = v_D \left( \kappa_x \sigma_x + \kappa_y \sigma_y \right) + \eta \left( \alpha s_z + \beta \right) \sigma_z - \Delta_c s_x, \qquad (1)$$

where $\sigma_i$ and $s_i$ are Pauli matrices that label the valley vortex pseudospin [49,56] and layer pseudospin, respectively, and $\left( \kappa_x, \kappa_y \right)$ is the wavevector deviated from the K point. In Eq. (1), the first term gives two overlapped conic dispersions with velocity $v_D$, the second term describes the bandgap opened by rotating rods, and the final term depicts the interlayer coupling that contributes to the frequency split. The parameter $\eta$ depends on the spatial filling ratio of the triangular rods, and $\Delta_c$ depends on the detailed geometry of the holes connecting the bilayers, both of which can be determined from numerical simulations. From the analytical model, one may derive a concise formula to characterize the curved phase boundary [Fig. 2(b)], i.e., $\beta^2 = \alpha^2 + \eta^{-2} \Delta_c^2$. Specifically, $\alpha = 0$ and $\beta_0 = \pm \Delta_c / \eta$ correspond to the critical angles where the nodal ring dispersions exist, which gives the straight phase boundary. The separated phase domains correspond to either nontrivial acoustic valley Hall (AVH) phases (shadowed regions) or nontrivial acoustic layer-valley Hall (ALH) phases (unshadowed regions), distinguished by the quantized topological invariant $C_V^K$ or $C_L^K$ (Supplemental Materials [56]). The former is a natural bilayer extension of the valley Chern number concerned in the monolayer system [48-54], and the latter identifies the layer information and resembles that proposed for QSH systems [57,58].

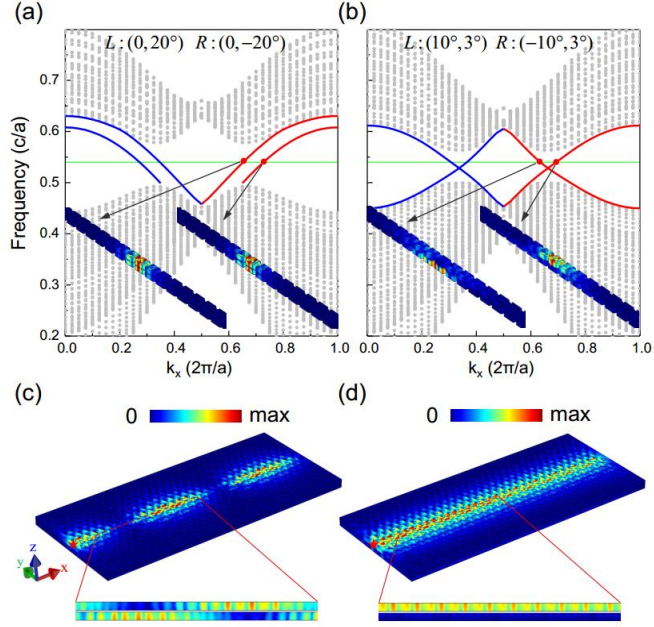

FIG. 3. Projected dispersions along an interface separating two topologically distinct AVH BSCs (a), and two topologically distinct ALH BSCs (b). Rod orientations are labeled for the BSCs located at the left (*L*) and right (*R*) sides. Red and blue curves correspond to the edge modes projected by the K and K' valleys, respectively. The insets exemplify two eigenstates projected to the K valley. (c) The amplitude field simulated for a finite-size sample with an AVH interface, excited by a point source (red star) positioned at the left entrance of the upper layer. A side view of the interface is zoomed-in to show the sound distribution in both layers. (d) The same as (c), but for the ALH case.

A smoking gun evidence for the topologically distinct phases is the presence of the edge modes. Figure 3(a) shows the numerical dispersion for an interface between two BSCs with rod orientations $(0°, \pm 20°)$, which belong to different AVH phases ($\Delta C_V^K = 2$). As anticipated from the bulk-boundary correspondence, two edge modes (red curves) carrying positive group velocities appear in the K valley, whose eigenfields disperse in both layers (see insets) and indicate a mixing of layer-pseudospins. In contrast, Fig. 3(b) shows the interface dispersion for a system constructed by two BSCs with rod orientations $(\pm 10°, 3°)$, which correspond to topologically distinct ALH phases ($\Delta C_L^K = 2$). Strikingly different from the AVH case, the two edge modes carry different group velocities. In particular, the eigenfield concentrates dominantly in either the upper layer or the lower layer. Therefore, the physics for the specific K valley resembles the QSH effect in electronic systems, i.e., the edge modes with different spins counter propagate. Because of the TR symmetry, accordingly, Figs. 3(a) and 3(b) also demonstrate a pair of edge modes (blue curves) in the K' valley, associated with opposite group velocities to those in the K valley. Critical signatures for the AVH and ALH systems are further demonstrated in Figs. 3(c) and 3(d), the sound propagations in finite-size samples stimulated by a point source. The point source is located at the left entrance of the upper layer, and thus excites only the edge modes that travel rightwards. It is observed that, the pressure field in Fig. 3(c) disperses in both layers: the strong and weak amplitudes vary alternatively in the upper and lower layers because of the interference between the two edge modes projected by the K valley, where the beat oscillation is determined by the momentum difference between the modes. The pressure field in Fig. 3(d), however, shows an excellent confinement in the upper layer. This indicates that, only the upper layer-polarized edge mode in the K valley is well excited, in contrast to the lower layer-polarized edge mode in the K' valley. The layer-selective excitation of the ALH edge modes could be particularly useful in real applications.

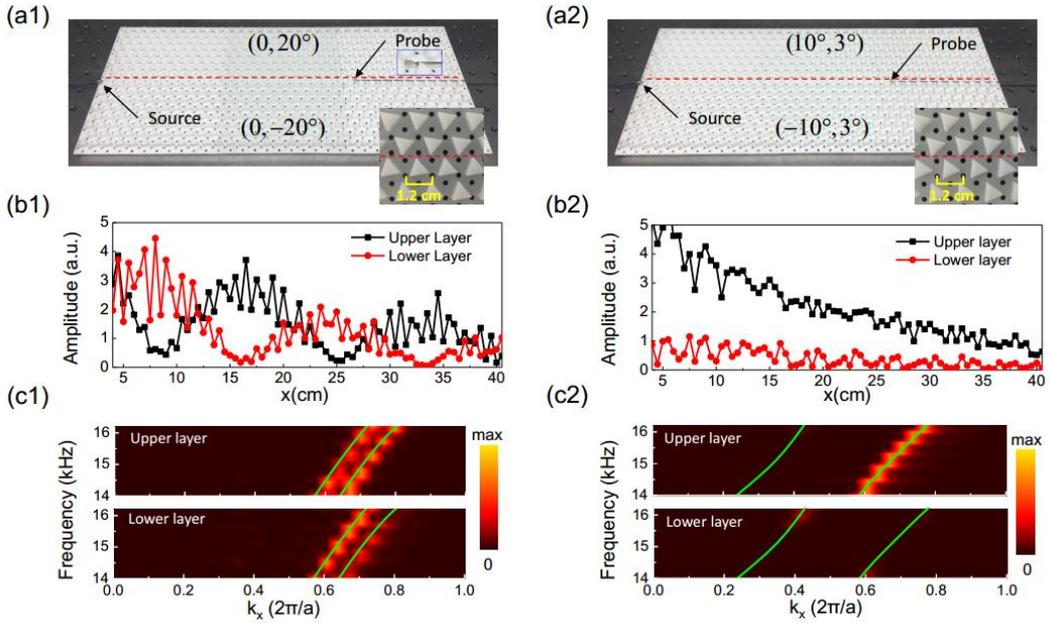

FIG. 4. Experimental evidences for the edge modes involved in Fig. 3. (a1) Experimental setup for the AVH case, where the red dashed line indicates the crystal interface. Inset: local view of the sample. The cover plate is removed for displaying more sample details. (b1) Experimental pressure amplitudes along the crystal interface, measured for both layers at 15.3kHz. (c1) Experimental edge dispersions

(bright color) achieved by Fourier transforming the pressure fields scanned inside the upper and lower layers, where the green curves label the simulated edge modes with positive group velocities. (a2)-(c2): The same as (a1)-(c1), but for the ALH case.

The topologically protected edge states for the AVH and ALH interface systems have been confirmed by experiments. Figures 4(a1) and 4(a2) demonstrate the experimental samples, where all geometric parameters are exactly those mentioned in Fig. 3 (with $a = 1.2$cm). Both samples have a size of 41cm x 18cm, made of 1056 triangular rods in total. The samples are prepared by three-dimensional printing, where the polymer material used can be safely viewed as acoustically rigid with respect to the air background. In our experiments, a sound generator is placed at the left entrance of the upper layer, and a subwavelength-sized sound probe is inserted inside the sample to scan sound profiles. We focus on the pressure distributions along the crystal interfaces, where the associated decay feature away from them has been checked. The measured pressure distributions for the AVH and ALH samples are presented in Figs. 4(b1) and 4(b2), respectively. As predicted above, the experimental results for the AVH sample manifest a clear beat effect associated with alternative wave concentrations in both layers, whereas the data for the ALH sample demonstrate a dominant field distribution in one layer. (The wave decay along the interface stems from the viscous dissipation.) Furthermore, the Fourier transforms of the pressure distributions give precise interface dispersions for the rightward propagating modes [Figs. 4(c1) and 4(c2)]. (The resolution in momentum can be refined by further lengthening the samples.) Again, the data capture well the layer-mixed and layer-polarized signatures respectively for the AVH and ALH edge modes: both AVH modes are excited in both layers, whereas the ALH modes are excited selectively.

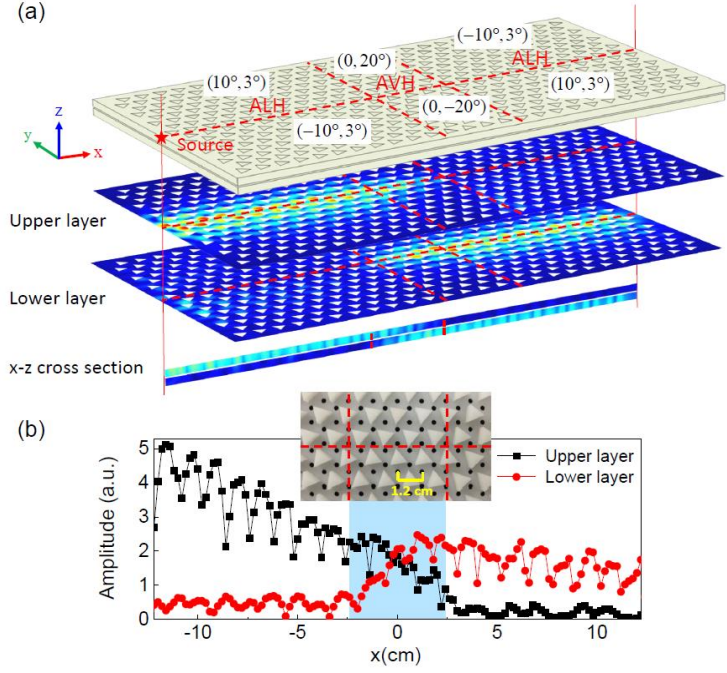

FIG. 5. A layer-polarization converter constructed by four BSCs with rod orientations $(\pm 10°, 3°)$ and $(0, \pm 20°)$, belonging to ALH and AVH acoustic insulators, respectively. (a) Full-wave simulation at 15.3kHz, which shows sound concentration varying from the upper layer to the lower layer. (b) Pressure amplitudes measured for the upper and lower interfaces along the $x$ direction, where the shadow region indicates clearly the inter-layer conversion.

Novel sound manipulations, e.g., intra/inter-layer communications, could be realized by integrating the above topological phenomena together in a compact device. Here we propose an efficient inter-layer converter to flip the layer-polarization. As shown in Fig. 5(a), the device is constructed by four distinct BSC phases that support ALH (bilateral) and AVH (middle) edge modes along the $x$ direction interfaces. For a point source (red star) positioned at the left entrance of the upper layer crystal interface, as predicted by a systemic analysis similar to Fig. 4, most of sound energy is switched to the lower layer as the wave reaches another ALH interface, assisted by the layer-mixed AVH interface with specific length. This inter-layer conversion has been further confirmed experimentally, as manifested in Fig. 5(b) by the measured pressure amplitudes along the upper and lower crystal interfaces.

In conclusion, a unique bilayer design of the SC has been proposed to attain and enrich topologically distinct acoustic insulators. Assisted with the additional layer information, the valley-projected edge modes can be either layer-mixed or layer-polarized. Interestingly, the bilayer design based on rotating scatterers is much different from those bilayer systems proposed in condensed matter physics [37-43], which allows us to explore fundamentally new physics beyond the original ones. Our findings have demonstrated versatile controllability over the valley-projected edge modes in response to external sound sources, comparing with the existing topological acoustic insulators [27-29,49]. Extensions of our scheme to other artificial structures (e.g., for elastic and electromagnetic waves) would be very interesting, and their further couplings with intrinsic polarizations may inspire both fundamental physics and practical applications.

**Acknowledgements**

This work is supported by the National Basic Research Program of China (Grant No. 2015CB755500); National Natural Science Foundation of China (Grant Nos. 11704128, 11774275, 11674250, 11534013, and 11747310); National Postdoctoral Program for Innovative Talents (BX201600054); China Postdoctoral Science Foundation (2017M610518).